\documentclass[manuscript]{aastex}

\usepackage{color}
\shorttitle{On the exoplanetary synchrotron radio bursts}
\shortauthors{Y. Gao et al.}

\begin{document}

%% LaTeX will automatically break titles if they run longer than
%% one line. However, you may use \\ to force a line break if
%% you desire.

\title{Observational features of exoplanetary synchrotron radio bursts}

%% Use \author, \affil, and the \and command to format
%% author and affiliation information.
%% Note that \email has replaced the old \authoremail command
%% from AASTeX v4.0. You can use \email to mark an email address
%% anywhere in the paper, not just in the front matter.
%% As in the title, use \\ to force line breaks.

\author{Yang Gao\altaffilmark{1}}
\author{Lei Qian\altaffilmark{2}}
\author{Di Li\altaffilmark{2, 3, 4}}
\affil{$^1$ School of Physics and Astronomy, Sun Yat-Sen University, Zhuhai 519082, Guangdong, People's Republic of China}
\affil{$^2$ CAS Key Laboratory of FAST, National Astronomical Observatories, Chinese Academy of Sciences, Beijing 100101, People's Republic of China}
\affil{$^3$ University of Chinese Academy of Sciences, Beijing 100049, People's Republic of China}
\affil{$^4$ NAOC-UKZN Computational Astrophysics Centre, University of KwaZulu-Natal, Durban 4000, South Africa}

%\author{C. D. Biemesderfer\altaffilmark{4,5}}
%\affil{National Optical Astronomy Observatories, Tucson, AZ 85719}

%\and

%\author{R. J. Hanisch\altaffilmark{5}}
%\affil{Space Telescope Science Institute, Baltimore, MD 21218}

%% Notice that each of these authors has alternate affiliations, which
%% are identified by the \altaffilmark after each name.  Specify alternate
%% affiliation information with \altaffiltext, with one command per each
%% affiliation.

%\altaffiltext{1}{here}

%% Mark off your abstract in the ``abstract'' environment. In the manuscript
%% style, abstract will output a Received/Accepted line after the
%% title and affiliation information. No date will appear since the author
%% does not have this information. The dates will be filled in by the
%% editorial office after submission.

\begin{abstract}
Magnetic fields of exoplanets are important in shielding the planets from cosmic rays and interplanetary plasma.
Due to the interaction with the electrons from their host stars,
  the exoplanetary magnetospheres are predicted to have both cyclotron and synchrotron radio emissions,
  of which neither has been definitely identified in observations yet.
As the coherent cyclotron emission has been extensively studied in literatures,
  here we focus on the planetary synchrotron radiation with bursty behaviors (i.e., radio flares) caused by the
  outbreaks of energetic electron ejections from the host star.
Two key parameters of the bursty synchrotron emissions, namely the flux density and burst rate,
  and two key features namely the burst light curve and frequency shift,
  are predicted for star - hot Jupiter systems.
The planetary orbital phase - burst rate relation is also considered as the signature of
  star-planet interactions (SPI).
As examples, previous X-ray and radio observations of two well studied candidate systems,
  HD 189733 and V830 $\tau$, are adopted
  to predict their specific burst rates and fluxes of bursty synchrotron emissions for further observational confirmations.
The detectability of such emissions by current and upcoming radio telescopes shows
  that we are at the dawn of discoveries.

%MEETING ABSTRACT
%We show here basic properties of relativistic combustion waves, namely detonation and deflagration waves, in scenes of astrophysical %processes. The Rankine-Hugoniot relations and the presences of detonation and deflagration waves are shown and analyzed in a general %relativistic case, which can be reduced to both non-relativistic and highly
%relativistic limits that are well established previously. Due to extreme conditions and various reaction mechanisms in astrophysics, %week detonation and strong deflagration, which are not valid in normal laboratorial conditions, are expected to exist in the %astrophysical environment. The relativistic theory of combustion waves is potentially applicable to the mechanism of gamma-ray burst %and the evolution of the early universe.
\end{abstract}

%% Keywords should appear after the \end{abstract} command. The uncommented
%% example has been keyed in ApJ style. See the instructions to authors
%% for the journal to which you are submitting your paper to determine
%% what keyword punctuation is appropriate.


%% From the front matter, we move on to the body of the paper.
%% In the first two sections, notice the use of the natbib \citep
%% and \citet commands to identify citations.  The citations are
%% tied to the reference list via symbolic KEYs. The KEY corresponds
%% to the KEY in the \bibitem in the reference list below. We have
%% chosen the first three characters of the first author's name plus
%% the last two numeral of the year of publication as our KEY for
%% each reference.

%% Authors who wish to have the most important objects in their paper
%% linked in the electronic edition to a data center may do so by tagging
%% their objects with \objectname{} or \object{}.  Each macro takes the
%% object name as its required argument. The optional, square-bracket
%% argument should be used in cases where the data center identification
%% differs from what is to be printed in the paper.  The text appearing
%% in curly braces is what will appear in print in the published paper.
%% If the object name is recognized by the data centers, it will be linked
%% in the electronic edition to the object data available at the data centers
%%
%% Note that for sources with brackets in their names, e.g. [WEG2004] 14h-090,
%% the brackets must be escaped with backslashes when used in the first
%% square-bracket argument, for instance, \object[\[WEG2004\] 14h-090]{90}).
%%  Otherwise, LaTeX will issue an error.

\section{Introduction}

As the key signal from exoplanetary magnetospheres,
  there have been efforts in detecting the radio emission from exoplanets since,
  and even before the discovery of the first exoplanet
  \citep{yantis1977,bastian2000,sirothia2014,lynch2018,route2019}.
Although radio detections are achieved for a few exoplanetary systems, no definite conclusion has been made about whether
  the radiations are from the planets, their host stars, or even other radio sources close to the targets
  \citep{sirothia2014,bower2016}.
The difficulty of detection and further confirmation of the exoplanetary radio emissions lies on three factors to be quantified:
  (I) the emission frequency and corresponding radio flux density,
  (II) the rate of bursty emissions bearing different energies,
  and (III) the light curve and possible temporal - frequency shift features of the signal.

Because of the expected high radio flux, efforts in the early champion focus on the detection of
  exoplanetary coherent cyclotron emissions
  \citep[also mentioned as electron cyclotron masers (ECM), cf.][]{wu1979,dulk1985}.
However, the upper frequency limit of the ECM is only $\sim40$ MHz for Jupiter and
  even lower for exoplanets with magnetic fields weaker than that of Jupiter.
Observations using the ground based low frequency ($<$ 10 MHz) radio telescopes experience the ionosphere absorption,
  making the detection of exoplanets very hard.
Efforts on this low frequency branch will eventually rely on the future Lunar low frequency radio telescope array.
The upcoming SKA, with significant increase of sensitivity, is also expected
  to detect exoplanetary ECM above a few 10 MHz \citep{zarka2015,pope2019}.
On the other hand, there are special systems from which higher frequency ECM is expected.
For exoplanets with magnetic fields much stronger than that of Jupiter \citep[][]{cauley2019},
  and white dwarf (WD) - terrestrial planet systems \citep{willes2005,vanderburg2015,manser2019},
  the ECM frequency can reach $\sim100$ MHz or even higher.
Such systems are potentially detectable by LOFAR, GMRT or other state-of-the-art telescopes.

Generally, when we shift to relatively higher radio frequencies of $\sim 100$ MHz to a few tens GHz,
  the planetary synchrotron emission caused by high speed (relativistic) electrons from the host star dominates the spectra.
Considering the well observed X-ray flares from a few exoplanetary systems
  as the high energy counterparts of their synchrotron emissions
  \citep{pillitteri2010,pillitteri2011,pillitteri2014,poppenhaeger2013,maggio2015},
  the detection of the synchrotron radio emissions is also expected.
However, without the amplification mechanism as in cyclotron masers, the synchrotron emission flux is 5 orders of magnitude
  lower than the ECM emission \citep[Jupiter as an example, cf.][]{zarka2015}.
Extensive observations of Jupiter synchrotron emissions
  well characterized their flux, bursty behavior and even the particle energy distribution
  \citep{depater2004,kloosterman2008,bhardwaj2009,lou2012,becker2017}.
Compared to Jupiter, there are four variations for exoplanetary systems that affect their synchrotron radiation:
  (1) the distance between the exoplanet and its host star,
  (2) the magnetic field strength of the exoplanet and the magnetic field structure of the star-planet system,
  (3) the Lorentz factor of particles from the host star,
  and (4) the bursty behavior in the host star.

By considering the above four variations,
   in this paper we calculate the expected synchrotron radio flux density based on the comparisons with Jupiter,
   and estimate the burst rate according to the results from stellar flares in Section 2;
   clarify the light curve and frequency shift as the key features of planetary synchrotron radio bursts in Section 3;
   make a case study for HD 189733 system and V 830 $\tau$ system in Section 4;
   and conclude and discuss the results in Section 5.

\section{The radio synchrotron flux and burst rate}

\begin{figure}[ht]
 \epsscale{0.7} \plotone{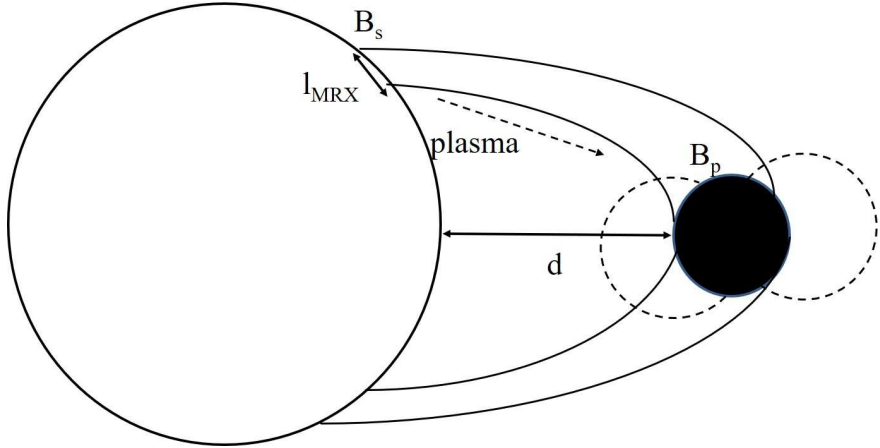}
\caption{Illustration of the magnetic field structure of a star - planet system close to each other.
  The left circle is the host star, linked by its magnetic field lines (solid curves) to the planet
  shown as the black filled circle.
  Magnetic fields of the planet are shown in dash lines, with its deformation when intercepting
  the stellar magnetic field not considered.
  Energetic particles arise from the active regions on the stellar surface and travel to the planet along the magnetic
  field loops.
  On the stellar surface, the active region where the magnetic reconnection (MRX) occurs has a typical scale of $l_{\rm MRX}$;
  and the average orbital radius of the planet is $d$.
  }
\end{figure}

For a hot Jupiter enclosed in the magnetosphere of its host star, the magnetic field structure,
  as well as the energetic particle production and transportation are schematically shown in Fig. 1.
Like in binary magnetized stars, radio active star - close planet systems are expected to have the main magnetic field loops
  originate from the host star and reach the planetary magnetic field \citep[cf.][]{uchida1983,dulk1985,trigilio2018,lanza2018}.
The average magnetic field strength at the stellar surface and planetary surface are noted as
  $B_{\rm s}$ and $B_{\rm p}$ respectively.

Energetic particles leading to synchrotron radiations are mainly from the coronal mass ejection (CME) in stellar surface
  through magnetic reconnection (MRX) processes \citep{zweibel2009,ji2011}.
In order to estimate the time duration of reconnection,
   we assume the MRX to occur within a site with the length scale $l_{\rm MRX}$.
The charged particles produced from the MRX site then travel along the magnetic field lines and reach the planetary
  magnetic field, leading to synchrotron emissions at initially the stellar surface and later also the planetary surface.
We would also indicate the average orbital radius of the planet $d$, which is important in the following calculations
  of the particle number at the planet, and the time-lag between the emissions from the star and the planet.

\subsection{Exoplanetary quiescent synchrotron radiation power}

The frequency of the synchrotron radiation from an electron with Lorentz factor $\gamma$ in magnetic field $B$ is
  \citep{rybicki1979,depater2004}
\begin{equation}
\nu\simeq1.3\gamma^2B \quad {\rm MHz},
\label{frequency1}
\end{equation}
  with $B$ in unit of Gauss.
Then for the detected electrons with energy $\sim$ 10 MeV in the Jupiter magnetic field of $\sim 4$ G,
  the synchrotron frequency is up to a few GHz, with the lower end extends to several tens MHz
  considering the variation of both
  $\gamma$ and $B$, which is consistent with observations.
It is also noted that as a result of both the energy distribution of relativistic electrons
  and the variation of magnetic field strength,
  the synchrotron radiation flux from Jupiter between $\sim 100$ MHz to a few GHz varies slightly within 3 to 5 Jy \citep{depater2004,kloosterman2008,bhardwaj2009,girard2016,becker2017},
  so we roughly consider it as a constant $\sim 4$ Jy.

Based on the synchrotron radiation power of a single electron,
  the total synchrotron radiation power can be written as
  \begin{equation}
  P=nP_{\rm e}\propto n\gamma^2 B^2,
  \label{power1}
  \end{equation}
  with $n$ the nominal electron number (without considering specific electron energy distributions).
It is readily seen that radiations from an exoplanet and its host star are different in power due to
  their different magnetic fields.
On the other hand, because the corresponding synchrotron radiation cooling time
  $\tau=8\times10^8 B^{-2} \gamma^{-1}\simeq2\times10^6$ s
  is much longer than the traveling time of a relativistic electron from the host
  star to the exoplanet in the case of a typical hot Jupiter ($t=0.1{\rm AU}/c=50$ s),
  the Lorentz factor of electrons can be considered as the same at the star and the planet.
At last we consider the variation of electron nominal numbers in (\ref{power1}) using the total electron numbers instead.
In quiescent synchrotron radiations caused by nearly isotropic stellar winds,
  electrons escape from the entire surface of the host star uniformly rather than from a single MRX site,
  so the structure in Fig. 1 does not apply.
Then the electron number density decreases as $r^{-2}$ as they travel away from the star,
  and the ratio of the numbers of electrons reach the planet $n_{\rm p}$ and radiate at the star $n_{\rm s}$
  can be calculated by simply considering the geometry, i.e.
\begin{equation}
  \frac{n_{\rm p}}{n_{\rm s}}=\frac{{R}^2}{4d^2},
  \label{nratio}
\end{equation}
  where $R$ is the radius of the planetary magnetosphere.
Then the overall ratio between the planetary radiation power and the stellar radiation power is
  \begin{equation}
  \frac{P_{\rm p}}{P_{\rm s}}=\frac{R^2}{4d^2}\frac{B_{\rm p}^2}{B_{\rm s}^2}.
  \label{pratio}
\end{equation}

For Jupiter, the estimated power ratio to the solar synchrotron radiation is of the order $0.5\times10^{-5}$
  (where we have adopted the Jupiter magnetic field of 4 G, magnetosphere radius of $\sim$ 20 times
  the Jupiter radius, and solar magnetic field of 2 G),
  which is consistent with observations \citep{depater2004,griebmeier2006}.
Then for a hot Jupiter with typical orbital radius $\sim 0.1$ AU and magnetic field $\sim 4$ G,
  if we still assume the magnetic field of its host star to be $\sim 2$ G,
  the power ratio is $\frac{P_{\rm p}}{P_{\rm s}}\sim 0.8\times 10^{-2}$,
  i.e., the quiescent radiation power from a hot Jupiter is about $\sim0.8\%$ of the quiescent radiation power from its host star.
Considering that (1) the hot Jupiter magnetic field could be weaker due to possible spin slow-down by tidal lock,
  (2) the planetary magnetosphere could be smaller in size as it is more compressed being closer to the host star,
  and (3) the host star magnetic field could be stronger in the K, M or T-Tauri stars which we are interested in,
  the above power ratio should be usually smaller than $0.8\%$.
Of course a larger value of this power ratio is also possible
  for star - planet systems where the planetary magnetic field is much stronger
  \citep[cf.][where the inferred magnetic field should be further confirmed]{cauley2019}.

\subsection{Exoplanetary bursty synchrotron radiation: flux density}

Solar bursts have been well observed and classified to several types,
  among which type IV bursts that originate from the synchrotron emission of energetic electrons ($\sim 10$ MeV)
  along the corona based magnetic loops are of particular interest for our work.
The energetic electrons escape from the star in the CME,
  which provides an enhanced energetic plasma flux on the planet compared to the quiescent solar wind,
  and leads to flares in X-ray and radio emissions.
The origin of such energetic plasma ejection, although depends on the specific local magnetic field configuration,
  is generally believed to be the reconnection of magnetic field lines \citep{isobe2005,zweibel2009,ji2011}.
Then it is natural that for young, late type stars of K and M types including T-Tauri stars,
  flares are observed to be more common and stronger because of the more active magnetic fields therein
  \citep{dulk1985,white1992,feigelson1994,suters1996,gudel2003,stelzer2007}.
It is also reasonable that the first detection of radio flares toward an exoplanetary system was made on
  a T-Tauri star V830 $\tau$ \citep{bower2016,donati2017};
  and that observation efforts have been made towards a closer K star - planet system HD189733 \citep{route2019}.

Considering the bursty radiation from a planet enclosed in the magnetosphere of its host star,
  the magnetic field structure shown in Fig. 1 is adopted.
By further assuming the MRX site as the source of energetic electrons,
  which travel along the magnetic field lines to the planet,
  we consider the situation that the number of electrons experiencing synchrotron radiation in the planet
  is equal to that in its host star.
Then the ratio of the synchrotron radiation power only depends on the magnetic fields, being
   \begin{equation}
  \frac{P_{\rm p}}{P_{\rm s}}\Big|_{\rm burst}=\frac{B_{\rm p}^2}{B_{\rm s}^2}.
  \label{pratio2}
\end{equation}
There are two processes that reduce the electron transportation rate from the stellar MRX site
  to the planetary magnetic field,
  namely the retaining of electrons at the host star, and the dissipation during the transportation.
Considering the existence of local coronal loops that do not reach the planet,
  part of the electrons produced in the MRX retain to the stellar coronal.
According to the observations of binary magnetized stars \citep[e.g., UX Arietis in][]{mutel1985},
  the host stellar `core' radiation flux density takes only 10\% to 20\% of the binary `halo' flux density,
  meaning that most electrons travel to the loop connecting the binary.
As the type of the host star (K0) and the distance of the binary star ($\sim 0.1$ AU) in UX Arietis
  are similar to the active star - hot Jupiter systems studied here,
  the ratio of electrons retain to the stellar coronal is also neglectable in our scaling analysis.
To estimate the dissipation of electrons during the transportation,
  we calculate the gyro-radius of typical electrons of 10 MeV at 10 G magnetic field, being
  $r_{\rm gyro}=\frac{\gamma m_{\rm e}c^2}{eB}=0.07$ AU, with $m_{\rm e}$ and $e$ the electron mass and charge.
This gyro-radius is comparable to the scale of the magnetic field loop of $\sim0.1$ AU,
  meaning that only in systems with magnetic field in the loops stronger than $\sim10$ G,
  electrons with energy smaller than 10 MeV can travel to the planet without much dissipation.
Such strong magnetic field in the loops connecting the planet is possible if
  we assume the magnetic field in the corona active regions of the host stars to be $\sim 100$ G \citep{mutel1985,dulk1985}.

However,
  there are situations that the magnetic field loop connecting the stellar CME and planet is not closed \citep[cf.][]{uchida1983},
  in which the efficiency of electron transportation is smaller than one.
This occurs when the coronal magnetic field loop is not accurately directed to the planet.
As the extreme situation, the closed loop CMEs with 100\% electron transportation efficiency considered here
  maximize the radiation power from the planet,
  and is most likely to be directly detected.

To estimate the bursty flux density, we again start from the Jupiter observations.
The Jupiter quiescent synchrotron flux density is $\sim 4$ Jy at 4 AU through frequencies
  $\sim 100$ MHz to a few GHz,
  which is $1.6\times 10^{-11}$ Jy when putting it to a typical exoplanetary system at $10$ pc from the earth.
Comparing the geometric factors in Equ. (\ref{pratio}) for Jupiter and for hot Jupiters at 0.1 AU from their host stars,
  we can estimate the quiescent radiation flux density of a hot Jupiter with the same magnetic field as Jupiter (4 G)
  and with its host star similar to the sun,
  i.e., $I_{\rm p}=(4{\rm AU}/0.1{\rm AU})^2\times 1.6\times 10^{-11}$ Jy $\approx 2.5 \times 10^{-8}$ Jy.
Isotropic electron ejection from the quiet host star has been assumed in above calculations;
   while in the closed loop magnetic field with CME induced bursts (Fig. 1),
   all energetic electrons from the MRX site travel to the planetary magnetosphere.
Then if we further assume the stellar bursty radiation power to be identical to its isotropic quiescent radiation power,
  by omitting the geometry factor $\frac{R^2}{4d^2}$ in (\ref{pratio}),
  the planetary radiation flux density in the burst state (\ref{pratio2}) is readily
  $I_{\rm p}|_{\rm burst}=4\times(\frac{0.1{\rm AU}}{20\times 7\times10^9{\rm cm}})^2\times 2.5 \times10^{-8}$ Jy $\approx$ 0.01 mJy.
In these calculations the flux density ratio $I_{\rm p}/I_{\rm s}$ is identical to the radiation power ratio
  $P_{\rm p}/P_{\rm s}$;
  and the magnetosphere radius of hot Jupiters are assumed to be identical to that of Jupiter, being 20 times the Jupiter radius.

The above upper limit of 0.01 mJy is for planets around solar-like stars,
  and achieved by assuming that all energetic electrons from the host star
  are transported to the planet in the burst state.
So the variation of the planetary magnetosphere radius does not change this flux density.
Another assumption is that for the host star,
  the bursty power is the same in strength as the quiescent power,
  which is valid for quite a number of radio flare stars \citep{dulk1985,griebmeier2006}.
In addition, the energies of flares from these active stars sometimes exceed
  those of solar flares by several orders of magnitudes,
  with their flux densities reaching a few to several tens mJy for K and M stars \citep{abada1996,gudel2003},
  and even tens to hundreds mJy for T-Tauri stars \citep{white1992,suters1996}.
In these stellar systems, the upper limit of planetary bursty radiation in the closed loop CME situation considered here
  can also reach $\sim 1$ mJy level or even higher regarding their magnetic field strength (cf. Equ. (\ref{pratio2})).
To be noted is that these estimations are based on the single electron $+$ nominal number density description,
  which is a scaling approximation for the realistic electron energy distribution.

\subsection{Exoplanetary bursty synchrotron radiation: burst rate}

Above scaling analysis has shown that radio flares from some exoplanets with host stars being
  K and M type stars at $\sim 10$ pc from the Earth,
  or T-Tauri stars at $\sim 100$ pc from the Earth are observable with their
  flux densities $\sim 1$ mJy.
Then from the observational point of view, the following question arises: what is the rate of such bursts to be expected?
Statistics from solar and stellar flares have shown that the number of flares observed in a certain epoch of time,
  i.e., the flare rate, decreases while the flare energy increases.
Specifically, for the flare rate $dN$ within the flare energy interval $[E, E+dE]$,
\begin{equation}
\frac{dN}{dE}=kE^{-\alpha},
\label{rate}
\end{equation}
  with $\alpha$ close to 2 but varies for different types of flares and in different types of stars \citep{crosby1993,audard2000,gudel2003}.
For magnetically active stars, the rates of the most energetic flares with their X-ray energy being around $10^{33}$ ergs
  or higher are around a few times per ten days \citep[cf. Fig. 2 in][]{audard2000}.
According to the X-ray - radio correlation of flares \citep{benz1994,benz2010},
  such flares are expected to have their radio flux counterparts of $\sim 10$ mJy
  if we assume the stars to be at a distance of $\sim10$ pc from the earth
  and by further assuming a flare duration of $\sim 10^4$ s \citep[cf.][]{pillitteri2014}.
Such flare rate and duration then lead to
  a detection rate of $1\%$ to $10\%$ per each single observation with an integration time much shorter compared to the flare duration.
We note that both a shorter flare duration and a lower flare rate will lead to a lower detection rate.
Contrarily, a longer observation integration time may lead to a higher detection rate,
  which will be taken into account for specific sources in Section 4.

As we have assumed that the sources of flares are energetic electrons from the MRX in the stellar coronal (cf. Fig. 1),
  it is also interesting to estimate the detection rate based on the MRX rate.
The typical timescale for MRX in the stellar corona within length scale $l_{\rm MRX}$ can be estimated as
  \begin{equation}
 \tau_{\rm MRX}=\frac{l_{\rm MRX}}{v_{\rm A}},
 \label{MRXtime}
  \end{equation}
  with $  v_{\rm A}$ the Alfv\'{e}n speed in the stellar corona.
It is noted that here the stellar synchrotron radiation time for a single accelerated particle,
   being $\sim l_{\rm MRX}/c$ with $c$ the speed of light,
   is much smaller compared to the reconnection time (\ref{MRXtime}).
Thus during a single MRX, the ratio of the bursty radiation time over the quiescent radiation time
  is identical to the MRX rate,
  \begin{equation}
  \frac{\tau_{\rm MRX}}{\tau_{\rm quiet}}=\frac{v_{\rm R}}{v_{\rm A}}=M_{\rm MRX},
  \label{MRXrate}
  \end{equation}
  where $v_{\rm R}$ is the reconnection inflow speed and
  $M_{\rm MRX}$ is the reconnection rate usually having a value between 0.01 and 0.1 \citep{ji2011}.
Although the MRX rate $M_{\rm MRX}$ describes a single reconnection,
  it is also representative of the overall bursty radiation time over the quiescent time,
  if we further assume that MRX occurs continuously.
This readily leads to a flare detection rate of 1\% - 10\% when the observation time is shorter compared to the burst duration,
  consistent with observations of the most energetic flares with $\sim 10^4$ s durations at the rate of a few times per ten days  \citep[][]{audard2000,pillitteri2014}.

\section{Expected observational features of the planetary radio burst}

Before applying the above scaling estimations of radio flux densities and burst rates to the observation of realistic systems,
  we examine the other two observational features, i.e., the light curve and the frequency shift
  in bursty radiations bearing both contributions from the star and the planet.
These features will help to clarify whether the bursty radiation is only from the star, or from both the star and the planet.

\subsection{The light curve}

If the bursty radiations we observe contain contributions from both the star and the planet,
  the total observed flux density is simply the addition of the two.
More specifically, if we consider the time-lag $\tau_{\rm lag}$ between the initiation of stellar radiation
  and the initiation of exoplanetary radiation,
  it is simply the time for the relativistic electrons to travel from the star to the planet, i.e.,
\begin{equation}
  \tau_{\rm lag}=\frac{d}{\beta_{\rm max} c},
  \label{lagtime}
\end{equation}
  where $\beta_{\rm max}$ is the maximum speed of electrons divided by the speed of light $c$.
Observationally this time-lag can be used to calculate the exoplanetary orbit radius:
  for a typical hot Jupiter with $d\simeq0.1$ AU, $\tau_{\rm lag}\simeq50$ s.
The time-lag here is also the time between the end of stellar radiation and the end of exoplanetary radiation.

In addition to the time-lag between radiations from the two sources,
  the rising of the exoplanetary radiation to its full power also costs time,
  which is the time between the electrons with the maximum speed and with the minimum speed
  (that leads to radiations in observable frequencies) reach the exoplanet.
We note this rising time as $\tau_{\rm shift}$ as the radio frequency shifts during the rising of the
  exoplanetary radiation.
This shift time also applies to the quenching process of the exoplanetary radiation.
In reference to the detection of 10 MeV electrons in Jupiter,
  we consider the synchrotron electrons with maximum energy $\sim 15$ MeV ($\gamma=30$)
  and minimum energy $\sim 1.5$ MeV ($\gamma=3$),
  corresponding to emission frequency 12 GHz $>\nu>$ 120 MHz in an exoplanet with $B_{\rm p}=10$ G
  (cf. Equ. (\ref{frequency1})).
It is then readily to give $\tau_{\rm shift}=3.5$ s by adopting the above maximum and minimum electron speeds
  of 0.9995$c$ and $0.94c$ respectively, and using $d=0.1$ AU.
The time for the stellar radiation to rise to its full power, on the other hand,
  can be neglected if we assume the distance between the MRX site and the stellar radiation zone is small compared to the
  planetary orbital radius.

\begin{figure}[ht]
 \epsscale{0.7} \plotone{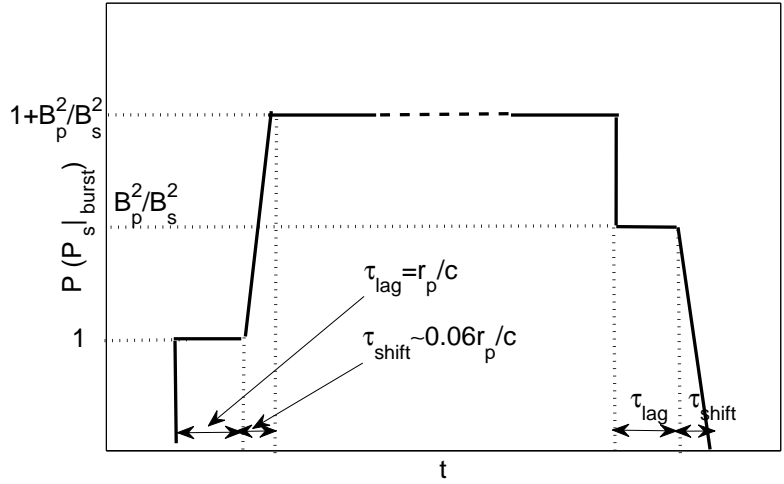}
\caption{Illustration of the three-stage bursty light curve of an exoplanet - host star system.
  The first stage is caused by the stellar radiation itself which lasts a period of $\tau_{\rm lag}=d/c$;
  followed by the full power with contributions from both the exoplanet and the host star with $1+B_{\rm p}^2/B_{\rm s}^2$
  times the power of the purely stellar emission;
  and ends with purely the exoplanetary radiation with power $B_{\rm p}^2/B_{\rm s}^2$ times the stellar emission,
  lasting for $\tau_{\rm lag}=d/c$.
  A shift time of $\tau_{\rm shift}\sim0.06d/c$ (varies with the observational frequency band and the planetary magnetic field) can also be noted due to the difference of times
    for electrons with different Lorentz factors to reach the exoplanet.}
\end{figure}

Both the lag time $\tau_{\rm lag}$ and shift time $\tau_{\rm shift}$ are schematically shown in the light curve in Fig. 2.
The magnitudes of the purely stellar (or exoplanetary) bursty radiation and the total bursty radiation,
  if both are well observed, can be used to estimate the ratio of the magnetic fields between the exoplanet and the host star.
This light curve is achieved under the assumption that stellar synchrotron radiation occurs within the MRX site.
Considering possible extensions of the length of the site producing observable radiations,
  the initial purely stellar radiation in Fig. 2 should not be a flat curve,
  but a rising curve instead.
Additionally, if specific electron energy distribution is considered,
  the purely stellar radiation should also be a rising curve instead of a flat.
However, an abrupt rise of the radiation power is still expected when the planetary radiation initiates.
In order to capture the rise or decay of the flare from exoplanetary systems,
  an observation time longer than or comparable to the burst duration is required.

\subsection{The frequency shift}

As electrons with different Lorentz factors reach the exoplanet at different times,
  during the rising time $\tau_{\rm shift}$ of the exoplanetary emission,
  the radiation frequency shifts from initially the maximum value to a band with the minimum value varying with time.
We first write the radiation frequency (\ref{frequency1}) in the form of the electron speed $\beta$:
\begin{equation}
\nu=1.3 B_{\rm p}\frac{1}{1-\beta^2}.
\label{frequency2}
\end{equation}
Then using the lag time (\ref{lagtime}) to express $\beta$ in the form of the electron traveling time
  from the star to the planet, we have $\beta=\frac{d}{ct}$.
By introducing this expression to (\ref{frequency2}) and rewrite the electron traveling time as $t=t_0+\tau$ with $t_0$
  the traveling time with the speed of light, the frequency depends on the time as
\begin{equation}
\nu=1.3 B_{\rm p} \frac{1}{1-\frac{d^2}{c^2(t_0+\tau)^2}}.
\label{frequency3}
\end{equation}
For relativistic electrons with the Lorentz factor greater than $\sim 2$,
  their speeds are smaller than the speed of light by $<10\%$;
  so we use the Taylor expansion in Equ. (\ref{frequency3}) for $\tau/t_0\ll1$ and get the simple expression as follows:
\begin{equation}
\nu= 1.3 B_{\rm p} \frac{t_0}{2\tau} = \frac{1.3 B_{\rm p} d}{2c}\frac{1}{\tau},
\label{frequencylinear}
\end{equation}
  or in the log-normal form:
\begin{equation}
{\rm ln}\nu={\rm ln}\big(\frac{1.3 B_{\rm p} d}{2c}\big) - {\rm ln}\tau.
\label{frequencylog}
\end{equation}

\begin{figure}[ht]
 \epsscale{0.7} \plotone{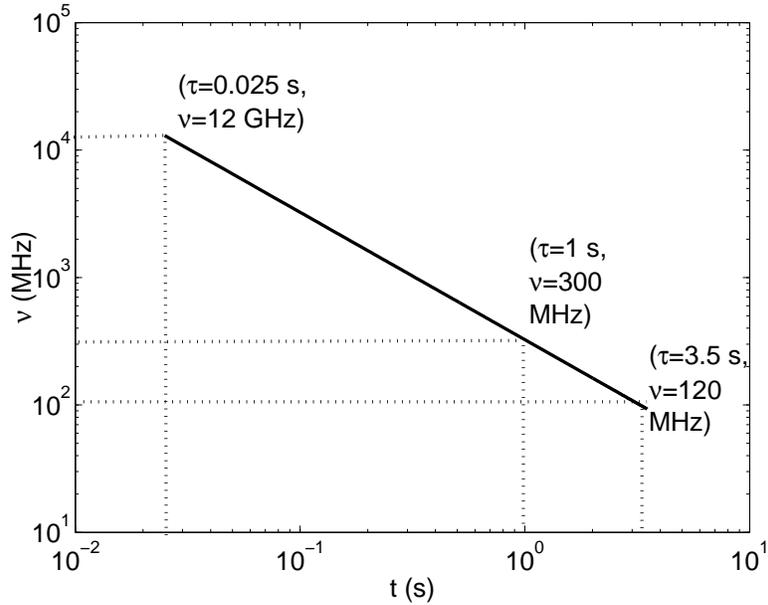}
\caption{Frequency shift at the rising and decaying periods of the exoplanetary radio burst.
  A shift of the emission band minimum frequency from 12 GHz to 120 MHz occurs within 3.5 s.
  Observationally to be noted is the frequency at $\tau=1$ s, which can be used to calculate the degenerated
  planet magnetic field $B_{\rm p}$ and orbit radius $d$ according to (\ref{frequencylog}).}
\end{figure}

As was indicated in the last subsection, the shift of frequency lasts about $\tau_{\rm shift}=3.5$ s
  to cover the range 12 GHz $>\nu>$ 120 MHz in an typical exoplanet with $B_{\rm p}=10$ G and $d=0.1$ AU.
It is then readily to plot the frequency (band minimum value) - time curve describing the frequency shift
  at the initiation and ending phases of the exoplanetary radiation as in Fig. 3.
According to Equ. (\ref{frequencylog}), the frequency at time $\tau=1$ s, if well measured in observations,
  can be used to measure the exoplanetary magnetic field $B_{\rm p}$,
  where the degeneracy with $d$ may be reduced by other observations (e.g., the time-lag $\tau_{\rm lag}$).
Above we calculated the continuous frequency shift at the rising and decaying periods of the exoplanetary radiation.
However, the key assumption in the calculation that electrons with different Lorentz factors are produced in MRX
  simultaneously, may not apply in realistic MRX particle accelerations.
More detailed calculations based on the time evolution of accelerated electron energy spectrum
  \citep[e.g.,][]{sironi2014,matsumoto2015} may lead to different results.

There is, however, another abrupt change of frequency if we pay attention to the difference of radiation frequency at
  the star and the planet for electrons with the same Lorentz factor $\gamma$.
In some exoplanetary systems,
  SPIs identified through Ca II K line activities indicate stronger planetary magnetic fields than in their host stars \citep{cauley2019},
  which is to be further confirmed in radio observations.
According to (\ref{frequency1}), an increase of the maximum frequency at the burst initiation phase is expected
  in this kind of systems.
Such frequency shift occurs at the typical time $\tau_{\rm lag}$, which is an order of magnitude longer then the
  frequency shift shown in Fig. 3, thus easier to be detected.
When the planetary magnetic field is weaker than that of the host star,
  in the ending epoch an corresponding decrease of the maximum frequency is similarly expected.
The abrupt shifts of the frequency then provide a measure of the magnetic field ratio $B_{\rm p}/B_{\rm s}$.
This frequency shift, accompanying the power changes as shown in Fig. 2,
  does not relay on the evolution of the electron energy spectral during MRX,
  thus is a more reliable method to confirm the existence of planetary radiation.
Examples of this kind of frequency shift due to the change of magnetic field can be found in type IV solar bursts,
  where the synchrotron frequency decreases as the energetic electrons travel out from the solar surface \citep{dulk1985};
  or S-bursts in the decameter emission from Jupiter \citep{clarke2014}.

\subsection{Orbital phase correlation}

The features of both the light curve (rising and decaying) and the frequency shift require a good time resolution
  to be observationally identified.
When a long integration time is required to achieve a good sensitivity for the detection of the bursts,
  which may last for $\sim 10^4$ s long,
  the planetary orbital phase - burst correlation can be used to identify whether the bursts have a star-planet interactions (SPI) origin.

As the stellar magnetospheres are usually not axisymmetric, both energetic plasma ejections
  and synchrotron emissions vary as the planet orbits around the host star.
Considering the non-axisymmetry of the energetic plasma ejections from the host star,
  the planetary bursty emission power and rate vary and should correlate with
  its orbital phase in reference to a fixed local longitude of the host star.
So in low-temporal-resolution observations, we suggest to plot the detection rate and/or flux density
  versus orbital phase diagram to help identifying the origin of bursts.
Besides the orbital phase correlation as the key practical feature of planetary emissions,
  the synchrotron burst emission is also distinct from thermal emissions for having recognizable circular polarizations \citep{dulk1985}.

\section{Case study: HD 189733 b and V830 $\tau$ b}

\subsection{HD 189733 b}

HD 189733 is a K (K1-K2) star - hot Jupiter system whose X-ray flares have been well observed by XMM-Newton,
  Swift and Chandra telescopes \citep{pillitteri2010,pillitteri2011,pillitteri2014,lecavelier2012,poppenhaeger2013}.
Observations at the optical and FUV bands, in particular the measurement of atomic lines such as the Ca II K line,
  show probable interaction between the star and its planet \citep{pillitteri2015},
  and indicate a very high planetary magnetic field of 20 - 50 G \citep{cauley2019}.
According to the measured X-ray flare flux and the traditional G\"{u}del-Benz relation,
  the radio flux of HD 189733 system can be estimated to be $0.01-0.09$ mJy,
  at a burst rate of $\sim 13\%$ with the typical burst duration of $\sim 8$ ks
  \citep{benz1994,benz2010,pillitteri2014}.
This estimated flux, although smaller compared to the estimation of \citet{route2019},
  may reach the 0.1 mJy level if the uncertainty in the index of the G\"{u}del-Benz relation is enlarged from 0.5 to 1.
Then it is not surprising that previous radio observations give a ``non-detection" result
  given their sensitivitis being larger than 0.3 mJy \citep[cf. the review in][]{route2019}.

For the expectation of upcoming observations, the sensitivity level of $\sim 0.1$ mJy can be achieved by
  a $\sim 100$-seconds integration with VLA at 4.5 GHz with 1 GHz bandwidth \citep{bower2016},
  or $\sim 300$-seconds integration with FAST at 1.4 GHz with 400 MHz bandwidth \citep{li2019}.
Achieving a sensitivity of a few $0.01$ mJy by increasing the integration time to the ks level is also possible.
However, to distinguish the HD 189733 emission from the Galactic background,
  the confusion limit of FAST \citep{zhang2018b} needs to be increased simultaneously to the $0.01$ mJy level
  by using a short baseline interferometer to reduce the main-beam-width to
  $\sim$ 10 arc-seconds level \citep[cf.][]{zarka2019}.
For the detection rate, given a burst rate of $13\%$ with the burst duration of $\sim 8$ ks,
  a single observation with integration time of 8 ks has $23 \%$ chance to capture the burst,
  if we consider that a $\sim$ 1 ks
  integration time is required to achieve the sensitivity of a few $0.01$ mJy.
Then a detectability of $93\%$ is expected in ten observations with each having 8 ks integration time.

However, such long integration time of kilo-seconds
  makes it impossible to temporally resolve the flux variation and frequency shift at $\tau_{\rm lag}=50$ s
  as shown in Section 3.
Such features can only be captured using the SKA2 for flares at 0.05 mJy level,
   or SKA1 for flares at 0.5 mJy level \citep[cf.][]{pope2019}.

Given the large number of nearby ($\sim 10$ pc) young, late type (K and M) stars with exoplanets discovered,
%  (cf. The Extrasolar Planets Encyclopaedia, http://www.exoplanet.eu),
  the detection of radio flares at a flux similar to the estimation of HD 189733 is expected
  on VLA and FAST (with reduced confusion limits).
Such observational attempts will be more efficient if proper selections from the sources
  with X-ray flares already detected are made \citep[e.g.,][]{maggio2015}.

\subsection{V830 $\tau$ b}

For the T Tauri star V830 $\tau$,
  the radio flare detection is even before the confirmation of the existence of a hot Jupiter around it
  \citep{donati2016,donati2017,bower2016}.
Detected by VLA and VLBI among the five observations in two separated epoches,
  the radio flux is $\sim 0.5$ mJy, with a burst rate of $\sim 40\%$ \citep{bower2016}.
Compared to HD 189733, an observational sensitivity at 0.1 mJy level
  (100-seconds VLA integration, or 300-seconds FAST integration without confusion from the background within the main-beam)
  should be good enough to detect such bursts.
Being lack of the information about the duration time of the bursts,
  we assume them to be much longer than the $\sim$300 s integration time of existing VLA observations;
  consequently a $40\%$ chance of detection for a single observation lasting 300 s is expected,
  and the detectability is $92\%$ in five observations.
To further capture the rising and ending light curves of a burst,
  a longer single observation time of a few kilo-seconds is required,
  consequently a higher detection rate is expected.

The flux change that occurs $\tau_{\rm lag}=50$ seconds after the burst initiation,
  as well as the accompanying abrupt frequency shift are expected to be observed
  by SKA1 at a sensitivity of 0.1 mJy for 5-seconds integration \citep{pope2019}.
The continuous frequency shift within $\tau_{\rm shift}\sim3.5$ s at the initiation of the planetary radiation
  is also expected to be resolved by SKA2 with its sensitivity better than SKA1 by an order of magnitude.

There are quite a number of T-Tauri stars with radio flare flux densities similar to V830 $\tau$ \citep{white1992,feigelson1994,suters1996}.
Although only a small portion of them have been detected bearing exoplanets due to the selection bias of current methods,
  the commonly existence of planets or protoplanets around T-Tauri stars is expected.
Then observations toward T-Tauri stars, as well as other K and M type stars with strong flares,
  will possibly detect the radio flare - (unknown) planetary orbital phase correlation,
  which may serve as a new method of exoplanet discovery.

\section{Conclusion and discussions}

Scaling analysis and applications to specific sources show that for the detection of exoplanetary synchrotron radio bursts
  from K and T-Tauri stars hosting planets,
  current telescopes VLA, FAST and Arecibo (with necessary interferometers to reduce the confusion limit for the later two)
  have the required sensitivity of a few $0.01$ mJy given enough integration time.
To resolve the signatures of both the flux variation and the frequency shift
  at the rising and ending processes of the burst (Fig. 2 and Section 3) in order to identify emissions from the exoplanet,
  SKA has the required sensitivity.
Before SKA, with the increasing number of radio detections from star-exoplanet systems,
  the orbital phase - burst correlation may serve as a prior way to discuss whether the
  radio emission is related with SPI \citep{pillitteri2014,maggio2015,route2019}.
For the selection of targets, systems with stronger host stellar magnetic field ($>\sim 100$ G)
  are expected to have higher electron transportation efficiency from the star to the planet,
  thus have stronger planetary synchrotron bursts.

M dwarfs and ultra cool dwarfs (UCD) are also radio active stars with their flare strength comparable to
  or stronger than K stars \citep{dulk1985,hughes2019}.
In the TESS era, the detection of hundreds of M dwarfs with optical flares
  provides a pool for further radio observations \citep{gunther2019,doyle2019}.
M dwarfs may host nearby planets either with strong magnetic fields,
  or without magnetic fields but electrically conductive.
The former systems are expected to have similar synchrotron radio bursts
  as was discussed in this paper;
  while the radiation features of flares from conductive exoplanets around M dwarfs is beyond the scope of this paper.

\acknowledgments
This work is supported by the National Natural Science Foundation of China grant No. 11988101, and by the startup fund from Sun Yat-Sen University.

\clearpage

\clearpage


\begin{thebibliography}{}



\bibitem[Abada-Simon (1996)]{abada1996}
    Abada-Simon, M. 1996, Plant. Space Sci., 44, 501
%Comparison of the observational data on flare stars, solar and planetary radio emissions

\bibitem[Audard et al.\ (2000)]{audard2000}
  Audard M, G\"{u}del, M., Drake, J. J. \& Kashyap, V. L. 2000, ApJ, 541, 396
% burst rate and duration

\bibitem[Bastian et al.\ (2000)]{bastian2000}
  Bastian, T. S., Dulk, G. A. \& Leblanc, Y. 2000, ApJ, 545, 1087

\bibitem[Becker et al.\ (2017)]{becker2017}
  Becker, H. N., Santos-Costa, D., J{\o}rgensen, J. L., Denver, T., Adriani A., et al. 2017, Geophys. Res. Lett., 44, 4481
%Observations of MeV electrons in Jupiter¡¯s innermost radiation belts and polar regions by the Juno radiation monitoring investigation: Perijoves 1 and 3

\bibitem[Benz \& G\"{u}del (1994)]{benz1994}
  Benz, A. O. \& G\"{u}del, M. 1994, A\&A, 285, 621
% Benz-Gudel relation

\bibitem[Benz \& G\"{u}del (2010)]{benz2010}
  Benz, A. O. \& G\"{u}del, M. 2010, ARA\&A, 48, 241
% Physical Processes in Magnetically Driven Flares on the Sun, Stars, and Young Stellar Objects

\bibitem[Bhardwaj et al.\ (2009)]{bhardwaj2009}
    Bhardwaj, A., Ishwara-Chandra, C. H., Shankar, N. U., Misawa, H., Imai, K. 2009, in \emph{The Low-Frequency Radio Universe}
    (Saikia, D. J., Green, D. A., Gupta, Y. and Venturi, T. eds., ASP Conference Series, Vol. 407, p.369-372)
%GMRT Observations of Jupiter¡¯s Synchrotron Radio Emission at 610 MHz

\bibitem[Bower et al.\ (2016)]{bower2016}
  Bower, G. C., Loinard, L., Dzib, S., Galli, P. A. B., Ortiz-Le\'{o}n, G. N., Moutou, C. \& Donati, J.-F. 2016, ApJ, 830, 107
%VARIABLE RADIO EMISSION FROM THE YOUNG STELLAR HOST OF A HOT JUPITER

\bibitem[Cauley et al.\ (2019)]{cauley2019}
Cauley, P. W., Shkolnik, E. L., Llama, J. \& Lanza, A. F. 2019, Nat. As., 3, 1128
%Magnetic field strengths of hot Jupiters from signals of star¨Cplanet interactions

\bibitem[Chadney et al.\ (2017)]{chadney2017}
    Chadney, J. M., Koskinen, T. T., Galand, M., Unruh, Y. C. \& Sanz-Forcada, J. 2017, A\&A 608, A75
% Effect of stellar flares on the upper atmospheres of HD 189733b and HD 209458b

\bibitem[Clarke et al.\ (2014)]{clarke2014}
  Clarke, T. E., Higgins, C. A., Skarda, J., Imai, K., Imai, M. et al. 2014, Journal of Geographysical Research: Space Physics, 119, 12
 %Probing Jovian decametric emission with the long wavelength array station 1

\bibitem[Crosby et al.\ (1993)]{crosby1993}
   Crosby, N. B., Aschwanden, M. J. \& Dennis, B. R. 1993, Solar Phys. 143, 275
% solar flare rate distribution

\bibitem[de Pater (2004)]{depater2004}
    de Pater, I. 2004, Planet. Space Sci., 52, 1449
% LOFAR and Jupiter¡¯s radio (synchrotron) emissions

\bibitem[Donati et al.\ (2016)]{donati2016}
  Donati, J. F., Moutou, C., Malo, L., Baruteau, C., Yu, L. et al. 2016, Nature, 534, 662
%A hot Jupiter orbiting a 2-million-year-old solar-mass T Tauri star

\bibitem[Donati et al.\ (2017)]{donati2017}
  Donati, J. F., Yu, L., Moutou, C., Camero, A. C., Malo, L. et al. 2017, MNRAS, 465, 3343
%The hot Jupiter of the magnetically active weak-line T Tauri star V830 Tau

\bibitem[Doyle et al.\ (2019)]{doyle2019}
Doyle, L., Ramsay, G., Doyle, J. G. \& Wu, K. 2019, MNRAS, 489, 437


\bibitem[Dulk (1985)]{dulk1985}
  Dulk, G. A. 1985, ARA\&A, 23, 169
%Radio Emission from the sun and stars

\bibitem[Feigelson et al.\ (1994)]{feigelson1994}
  Feigelson, E. D., Welty, A. D., Imhoff, C., Hall, J. C., Etzel, P. B., Phillips, R. B. \& Lonsdale, C. J. 1994, ApJ, 432, 373
%Multiwavelength Study of the Magnetically Active T Tauri Star HD 283447

\bibitem[Girard et al.\ (2016)]{girard2016}
  Girard, J. N., Zarka, P., Tasse, C., Hess, S., de Pater, I. et al. 2016, A\&A, 587, A3

\bibitem[Grie{\ss}meier (2006)]{griebmeier2006}
    Grie{\ss}meier, J.-M. 2006, Doktors der Naturwissenschaften,
%Aspects of the magnetosphere-stellar wind interaction of close-in extrasolar planets

\bibitem[Grie{\ss}meier (2016)]{griebmeier2016}
    Grie{\ss}meier, J.-M. 2016, in {Planetary Radio Emissions VIII, Proceedings of the 8th International Workshop} (Fischer, G., Mann, G., Panchenko, M., and Zarka, P. eds. Austrian Academy of Sciences Press, Vienna, 2017, p. 285-299)

\bibitem[G\"{u}del et al.\ (2003)]{gudel2003}
  G\"{u}del, M., Audard M., Drake J. J., Kashyap, V. L. \& Guinan, E. F.  2003, ApJ, 582, 423
% burst rate and duration

\bibitem[G\"{u}nther et al.\ (2020) ]{gunther2019}
G\"{u}nther, M. N., Zhan, Z., Seager, S. et al. 2019, AJ, 159, 60

\bibitem[Hughes et al.\ (2019)]{hughes2019}
    Hughes, A. G., Boley, A. C., Osten, R. A. \& White J. A. 2019, ApJ, 881, 33
%Constraining the Radio Emission of TRAPPIST-1

\bibitem[Isobe et al.\ (2005)]{isobe2005}
Isobe, H., Takasaki, H. \& Shibata, K. 2005, ApJ, 632, 1184
% magnetic reconnection rate in solar flares


\bibitem[Ji \& Daughton (2011)]{ji2011}
Ji, H. \& Daughton, W. 2011, Phys. Plasmas, 18, 111207



\bibitem[Kloosterman et al.\ (2008)]{kloosterman2008}
  Kloosterman, J. L., Butler, B. \& de Pater, I. 2008, Icarus, 193, 644
%VLA observations of synchrotron radiation at 15 GHz

\bibitem[Lanza (2018)]{lanza2018}
    Lanza, A. F. 2018, A\&A, 610, A81
%Close-by planets and flares in their host stars

\bibitem[Lecavelier des Etangs et al.\ (2012)]{lecavelier2012}
    Lecavelier des Etangs, A., Bourrier, V., Wheatley, P. J., Dupuy, H., Ehrenreich, D., et al., 2012, A\&A, 543, L4
% Temporal variations in the evaporating atmosphere of the exoplanet HD189733b

\bibitem[Li et al.\ (2019)]{li2019}
  Li, D., Dickey, J. M., \& Liu, S. 2019, RAA, 19, 16

\bibitem[Lou et al.\ (2012)]{lou2012}
  Lou, Y.-Q., Song, H., Liu, Y. \& Yang, M. 2012, MNRAS, 421, 62

\bibitem[Lynch et al.\ (2018)]{lynch2018}
    Lynch, C. R., Murphy, T., Lenc, E. \& Kaplan, D. L. 2018, MNRAS, 478-1763
%The detectability of radio emission from exoplanets

\bibitem[Maggio et al.\ (2015)]{maggio2015}
    Maggio, A., Pillitteri, I., Scandariato, G., Lanza, A. F., Sciortino, S. et al. 2015, ApJL, 811, L2
%COORDINATED X-RAY AND OPTICAL OBSERVATIONS OF STAR¨CPLANET INTERACTION IN HD 17156

\bibitem[Manser et al.\ (2019)]{manser2019}
   Manser, C. J.,  G\"{a}nsicke, B. T., Eggl, S. et al. 2019, Science, 364, 66

\bibitem[Matsumoto et al.\ (2015)]{matsumoto2015}
Matsumoto, Y., Amano, T., Kato, T. N. \& Hoshino, M. 2015, Science, 347, 974

\bibitem[Mutel et al.\ (1985)]{mutel1985}
Mutel, R. L., Lestrade, J. F., Preston, R. A. \& Phillips, R. B. 1985, ApJ, 289, 262

\bibitem[Pillitteri et al.\ (2010)]{pillitteri2010}
  Pillitteri, I., Wolk, S. J., Cohen, O., Kashyap, V., Knutson, H., Lisse, C. M. \& Henry, G. W. 2010, ApJ, 722, 1216
%XMM-NEWTON OBSERVATIONS OF HD 189733 DURING PLANETARY TRANSITS

\bibitem[Pillitteri et al.\ (2011)]{pillitteri2011}
  Pillitteri, I., G\"{u}nther, H. M., Wolk, S. J.,, Kashyap, V. L. \& Cohen, O. 2011, ApJL, 741, L81
%X-RAY ACTIVITY PHASED WITH PLANET MOTION IN HD 189733?

\bibitem[Pillitteri et al.\ (2014)]{pillitteri2014}
  Pillitteri, I., Wolk, S. J., Lopez-Santiago, J., G\"{u}nther, H. M., Sciortino, S., Cohen, O., Kashyap, V., \& Drake, J. J. 2014, ApJ, 785, 145
%THE CORONA OF HD 189733 AND ITS X-RAY ACTIVIT

\bibitem[Pillitteri et al.\ (2015)]{pillitteri2015}
  Pillitteri, I., Maggio, A., Micela, G., Sciortino, S., Wolk, S. J. \& Matsakos, T. 2015, ApJ, 805, 52
%FUV VARIABILITY OF HD 189733. IS THE STAR ACCRETING MATERIAL FROM ITS HOT JUPITER?

\bibitem[Pope et al.\ (2019)]{pope2019}
  Pope, B. J. S., Withers, P., Callingham, J. R. \& Vogt, M. F. 2019, MNRAS, 484, 648
% Exoplanet Transits with Next-Generation Radio Telescopes

\bibitem[Poppenhaeger et al.\ (2013)]{poppenhaeger2013}
  Poppenhaeger, K., Schmitt, J. H. M. M., \& Wolk, S. J. 2013, ApJ, 773, 62
%TRANSIT OBSERVATIONS OF THE HOT JUPITER HD 189733b AT X-RAY WAVELENGTHS

\bibitem[Route (2019)]{route2019}
  Route, M. 2019, ApJ, 872, 79
%The Rise of ROME. I. A Multiwavelength Analysis of the Star¨CPlanet Interaction in the HD 189733 System

\bibitem[Reiners \& Christensen (2010)]{reiners2010}
    Reiners, A. \& Christensen, U. R. 2010, A\&A, 522, A13
%A magnetic field evolution scenario for brown dwarfs and giant planets

\bibitem[Rybicki \& Lightman (1979)]{rybicki1979}
  Rybicki, G. B. \& Lightman A. P. 1979, Radiative Processes in Astrophysics, John Wiley \& Sons

\bibitem[Sirothia et al.\ (2014)]{sirothia2014}
   Sirothia, S. K., Lecavelier des Etangs, A., Gopal-Krishna, Kantharia, N. G., \& Ishwar-Chandra, C. H. 2014, A\&A 562, A108
%Search for 150 MHz radio emission from extrasolar planets in the TIFR GMRT Sky Survey

\bibitem[Sironi \& Spitkovsky (2014)]{sironi2014}
  Sironi, L. \& Spitkovsky, A. 2014, ApJL, 783, L21

\bibitem[Stelzer et al.\ (2007)]{stelzer2007}
  Stelzer, B., Flaccomio, E., Briggs, K. et al. 2007, A\&A, 468, 463
% TT star statistics

\bibitem[Suters et al.\ (1996)]{suters1996}
    Suters, M., Stewart, R. T., Brown, A. \& Zealey, W. 1996, ApJ, 111, 320
%VARIABLE RADIO SOURCES IN THE CORONA AUSTRALIS CLOUD

\bibitem[Trigilio et al.\ (2018)]{trigilio2018}
   Trigilio, C., Umana, G., Cavallaro, F., Agliozzo, C., Leto P., et al. 2018, MNRAS, 481, 217
% Detection of a Centauri at radio wavelengths: chromospheric emission and search for star-planet interaction

\bibitem[Uchida \& Sakurai (1983)]{uchida1983}
    Uchida, Y. \& Sakurai, T. 1983, in P. B. Byrne \& M. Ronodo (eds.), {\it Activaty of Red Dwarf Stars}, p. 629
% Magnetospheric Radio Emissions from Exoplanets with the SKA

\bibitem[Vanderburg et al.\ (2015)]{vanderburg2015}
  Vanderburg, A., Johnson, J. A., Rappaport, S. et al. 2015, Nature, 526, 546

\bibitem[Vidotto \& Donati (2017)]{vidotto2017}
  Vidotto, A. A. \& Donati, J.-F. 2017, A\&A, 602, A39
%Predicting radio emission from the newborn hot Jupiter V830 Tauri b and its host star
% Low frequency

\bibitem[Wang \& Loeb (2019)]{wang2019}
  Wang, X. \& Loeb A. 2019, ApJ, 874L, 23
%Nonthermal Emission from the Interaction of Magnetized Exoplanets with the Wind of Their Host Star

\bibitem[White et al.\ (1992)]{white1992}
  White, S. M., Pallavicini, R. \& Kundu, M. R. 1992, A\&A, 257, 557
% Radio survey of post and T Tauri stars

\bibitem[Willes \& Wu (2005)]{willes2005}
  Willes, A. J. \& Wu, K. 2005, A\&A, 432, 1091

\bibitem[Wu \& Lee (1979)]{wu1979}
  Wu, C. S. \& Lee, L. C. 1979, ApJ, 230, 621

\bibitem[Yamada et al.\ (2010)]{yamada2010}
Yamada, M., Kulsrud, R. \& Ji, H. 2010, Rev. Mod. Phys. 82, 603
% Magnetic reconnection review

\bibitem[Yantis et al.\ (1977)]{yantis1977}
  Yantis, W. F., Sullivan, III., W. T. \& Erickson, W. C. 1977, Bull Am. Astron. Soc., 9, 453

\bibitem[Zarka et al.\ (2015)]{zarka2015}
    Zarka, P., Lazio, T. J. W., Hallinan, G. \& "Cradle of Life" WG 2015, in {\it Advancing Astrophysics with the Square Kilometre Array} (Proceedings of Science (AASKA14) 120)
% Magnetospheric Radio Emissions from Exoplanets with the SKA

\bibitem[Zarka et al.\ (2019)]{zarka2019}
    Zarka, P., Li, D., Grie{\ss}meier, J.-M., Lamy, L., Girard, J. N., Hess, S. L. G., Lazio, T. J. W. \& Hallinan, G. 2019, RAA, 19, 23
% FAST

\bibitem[Zhang et al.\ (2018)]{zhang2018}
  Zhang, H., Gao, Y. \& Law, C. K. 2018, ApJ, 864, 167
  
\bibitem[Zhang et al.\ (2018b)]{zhang2018b}
  Zhang, Z.~B., Chandra, P., Huang, Y.~F. \& Li, D. 2018, ApJ, 865, 82

\bibitem[Zweibel \& Yamada (2009)]{zweibel2009}
Zweibel, E. G. \& Yamada, M. 2009, ARA\&A, 47, 291

\end{thebibliography}
\end{document}